\documentclass[aps, twocolumn, showpacs, preprintnumbers, amsmath, amssymb, prb]{revtex4-2}
\usepackage{multirow} % for cmd 'multirow', 'multicolumn'
\usepackage{tabularx}
\newcolumntype{Y}{>{\centering\arraybackslash}X}
\usepackage{tikz}
\usepackage{amsfonts}
\usepackage{amstext}
\usepackage{amsmath}
\usepackage{amssymb}
\usepackage{comment}
\usepackage{latexsym}
\usepackage{graphicx,epstopdf}
\usepackage{ulem}
\usepackage[colorlinks,linkcolor=blue,anchorcolor=blue,citecolor=blue]%
{hyperref}
\usepackage{times}
\usepackage{subfigure}
\usepackage{color}
\usepackage{dcolumn}
\usepackage{graphicx}
\usepackage{soul}
\usepackage{bbding}
\usepackage{xcolor}
\begin{document}

\title{Odd-even parity dependent transport in an annular Kitaev chain}
\date{\today }
\author{Wei Wang$^{1}$}
%\affiliation{School of Electronic, Electrical and Communication Engineering, University of Chinese Academy of Sciences, Beijing 100049, China}
\author{Zhen-Gang Zhu$^{1,2}$}
\email{zgzhu@ucas.ac.cn}
\author{Gang Su$^{3,4}$}
\email{gsu@ucas.ac.cn}
%\affiliation{School of Electronic, Electrical and Communication Engineering, University of Chinese Academy of Sciences, Beijing 100049, China}
\affiliation{
$^{1}$ School of Electronic, Electrical and Communication Engineering, University of Chinese Academy of Sciences, Beijing 100049, China.\\
$^{2}$ School of Physical Sciences, University of Chinese Academy of Sciences, Beijing 100049, China. \\
$^{3}$  Institute of Theoretical Physics, Chinese Academy of Sciences, Beijing 100190, China.\\
$^{4}$ Kavli Institute for Theoretical Sciences, University of Chinese Academy of Sciences, Beijing 100190, China.
 }

\begin{abstract}
We investigate the impact of magnetic flux and the odd-even parity of lattice points $N$ on electron transport in an annular Kitaev chain, with an explanation provided from the energy band perspective. Three transport mechanisms including direct transmission (DT), local Andreev reflection (LAR) and crossed Andreev reflection (CAR) are considered. In particular, the connection configuration of electrodes to different lattice sites is studied, where the case that the two electrodes connected to the sites are aligned along a diameter is called as symmetric connection and otherwise as asymmetric connections. For even $N$ and asymmetric connection, the vanished LAR and CAR in symmetric connection will emerge as peaks. A more prominent observation is that the symmetry of the two resonant peaks due to DT processes located at $\Phi = N\pi/3$ and $\Phi = 2N\pi/3$ for symmetric connection will be broken, and the peak at $\Phi = N\pi/3$ will be largely reduced,  where $\Phi$ is the magnetic flux. Moreover, the peaks around $\Phi = N\pi/3$ due to LAR and CAR processes grows drastically even larger than that from DT. For LAR and CAR processes, there is no peak around $\Phi = 2N\pi/3$ and transmission due to these two processes are completely suppressed for $\Phi>N\pi/2$. Moreover, it is found that the energy bands vary with $\Phi$ in a period of $N\pi$ and $2N\pi$ for even or odd $N$. We finally systematically analyze the influence of weak disorder on transport and demonstrate that these parity-dependent effects are robust in the presence of disorder. These behaviors reflect a complicated competition from DT, LAR and CAR processes and the parity of the lattice number in the Kitaev ring, which will be interested for the quantum device based on Kitaev chain.
\end{abstract}

\maketitle

\section{INTRODUCTION}\label{SC1}

 Topological superconductors are widely regarded as a pivotal platform for fault-tolerant topological quantum computation \cite{Aghaee2025,Hu2025}, primarily due to their ability to host Majorana zero modes (MZMs) \cite{Mourik2012,Deng2016,Jeon2017,Albrecht2016,Kong2019,Kong2021,Liu2024,Wang2015}. These exotic quasiparticles obey non-Abelian statistics \cite{Read2000,Alicea2011,Yu2023,Halperin2012,Zhou2019,Kim2026}, making them ideal building blocks for topologically protected qubits. The Kitaev chain \cite{Kitaev2001}, as the prototypical one-dimensional model of a topological superconductor, not only provides a transparent theoretical framework for understanding the physical nature of MZMs \cite{Dutta2017,Fu2008,Lutchyn2010,Jiang2011,Stern2019,Pedrocchi2012,Rahmani2019,Hua2019} but has also been experimentally emulated in engineered quantum systems such as semiconductor-superconductor hybrid nano-wires \cite{Bordin2024,Bordin2025,Wang2022,Dvir2023}. Recently, growing attention has been directed toward realizing the Kitaev model on a ring geometry-namely, an annular Kitaev chain with periodic boundary conditions \cite{Miao2022,Zhou2018,Liu2023,Zeng2016}, In this setup, threading a magnetic flux $\Phi $ through the ring enables continuous tuning across topological phase transitions \cite{Lucignano2013}, thereby providing a novel means to probe bulk topological properties without relying on edge states.

However, a key yet insufficiently explored question is how the odd-even parity of the total number of lattice sites $ N $ influences the transport characteristics of such a ring. In this annular system, the parity of $ N $ fundamentally determines the symmetry of the energy band structure and the magnetic-flux periodicity-specifically, $ N\pi $ for even $ N $ and $ 2N\pi $ for odd $ N $, and profoundly affects whether the superconducting energy gap opens or closes at specific flux values. In this work, by combining the non-equilibrium Green's function formalism with the Landauer-B{\"u}ttiker transport theory, we systematically investigate the responses of three distinct transport channels-DT, LAR, and CAR - under varying $ N $ and electrode configurations. We find that for even $N $ with symmetric electrode placement (e.g., diametrically opposed leads), LAR and CAR are nearly completely suppressed, while DT exhibits two symmetric resonant peaks at $ \Phi = N\pi/3 $ and $ \Phi = 2N\pi/3 $. In stark contrast, for odd $ N $, this symmetry is broken: the DT peak at $\Phi = N\pi/3 $ is strongly suppressed, whereas pronounced peaks emerge in LAR and CAR at the same flux; meanwhile, only the DT peak persists at $ \Phi = 2N\pi/3 $. Energy-band analysis reveals that this parity-dependent behavior stems from the magnetic-flux-controlled opening and closing of the superconducting gap: for even $ N $, the gap closes at both $ N\pi/3 $ and $ 2N\pi/3 $, enabling coherent electron transmission; for odd $ N $, the gap remains open at $ N\pi/3 $ but closes at $ 2N\pi/3 $, thereby favoring Andreev processes over direct transmission at the former. Moreover, this parity-sensitive transport signature remains robust against weak disorder, suggesting its experimental accessibility.

Our work demonstrates that the lattice parity acts as an intrinsic symmetry degree of freedom that can dramatically reshape quantum transport pathways in topological superconducting rings. This finding deepens the understanding of topological properties in finite-sized Kitaev systems and may provide a practical route for detecting topological phases through parity-resolved conductance measurements in future quantum devices.

The remainder of this paper is structured as follows. In Sec. \ref{SC2}, the theoretical model and the tight-binding Hamiltonian of the annular Kitaev chain are introduced. Here, we define the key parameters and interactions within the model, which serve as the foundation for subsequent analysis. In Sec. \ref{SC3}, the nonequilibrium Green's function method in conjunction with the Landauer-B{\"u}ttiker formula is employed to investigate the transmission coefficient of the model under the conditions of odd and even lattice points. Through a series of calculations and derivations, we analyze the intricate relationships between the lattice-point parity, the electronic states, and the transmission characteristics. This allows for an in-depth understanding of how the transport properties of the annular Kitaev chain are affected by $N$. Finally, in Sec. \ref{SC4}, a concise summary is provided.

\begin{figure}[tb]
\centering
\includegraphics[width=0.5\textwidth]{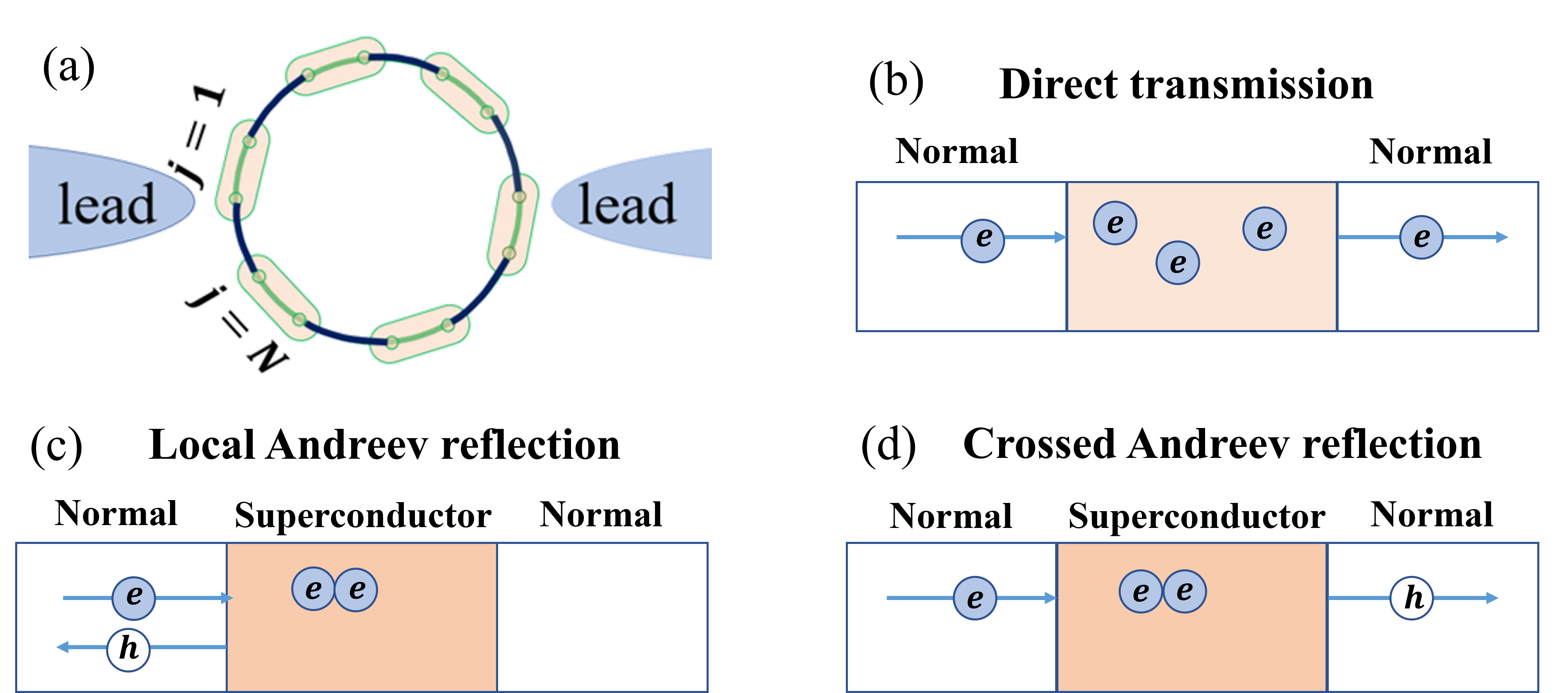}
\caption{(a) Schematic plot of the annular Kitaev chain model. The central region represents an annular Kitaev chain. Each ellipse denotes a site on the chain, and there are two MZMs at each site. The dark blue lines in the figure indicate the particle interactions between different sites, while the green lines represent the particle interactions at the same site. Electrodes are connected at the left and right ends of the chain. (b) Schematic illustration of DT. In this case, electrons are directly transmitted through the system, and the charge carriers are electrons. (c) Schematic illustration of local Andreev reflection (LAR). A single incoming electron being reflected as a hole, with a corresponding Cooper pair added to the superconductor. (d) Schematic illustration of crossed Andreev reflection (CAR). A single electron impinges on the superconductor interface, a hole is reflected from the opposite interface, with a Cooper pair enters into the superconductor.}
\label{Fig1}
\end{figure}

\section{MODEL AND METHODS}\label{SC2}

In this work, the number of lattice sites $N$ is changed in the ring structure. A schematic illustration of the model is presented in Fig. \ref{Fig1}(a), and an experimental approach is proposed based on Reference \cite{Bordin2024,Bordin2025}, this device consists of an InSb nano loops placed on top of an array of $4N$ finger gates separated by a thin dielectric. $N$ superconducting Al contacts are evaporated on top of the nanowire using the shadow-wall lithography technique. Both sides of the device are further contacted by two normal Cr/Au probes. Every contact is connected to an independent voltage source and current meter. The $N$ finger gates underneath the semiconductor-superconductor hybrid segments control their chemical potential, while the other $3N$ gates form quantum dots on each of the $N$ bare InSb sections. In this work, we give a toy model, which breaks the U(1) symmetry, and will simply assume that only one spin component is present in the electron spectrum \cite{Kitaev2001}. For the annular Kitaev chain in the central region, the tight-binding Hamiltonian is \cite{Miao2022}
\begin{eqnarray}
%  \begin{aligned}
H_\mathrm{C} &=& \sum_{j=1}^{N}-\mu c_{j}^{\dagger} c_{j} + \sum_{j=1}^{N-1}\left(-t e^{i\phi} c_{j}^{\dagger} c_{j+1}+|\Delta| e^{i\theta}e^{i\phi} c_{j}^{\dagger}c_{j+1}^{\dagger}\right) \notag\\
 & + & \left(-t e^{i\phi} c_{N}^{\dagger} c_{1}+|\Delta|e^{i\theta}e^{i\phi} c_{N}^{\dagger}c_{1}^{\dagger}\right)+\text{H.C.},
 \label{Eq1}
%  \end{aligned}
\end{eqnarray}
where $c_{j}^{\dagger}(c_{j})$ denotes the creation (annihilation) operator corresponding to lattice site $j$. The first term in the Hamiltonian represents the on-site potential energy term, with $\mu$ being the potential energy.
The second term describes the hopping energy and the pairing interaction between nearest-neighbor lattice sites along the Kitaev chain. $t$ is the nearest-neighbor hopping energy, and $\phi$ is the phase acquired due to the magnetic flux $\Phi$ threading the ring via the Peierls substitution, and $\Delta = |\Delta| e^{i\theta}$ the proximity-induced superconducting pairing gap.
The last term accounts for the hopping and the pairing interaction between the first and the last sites of the chain, signifying the coupling between lattice site $j = 1$ and lattice site $j = N$.
It is well-established that when $t = |\Delta|$ and $\mu = 0$, MZMs will emerge at both ends of the Kitaev chain \cite{Kitaev2001}.

In the physical realization, the magnetic flux $\Phi$ threading the ring is incorporated into the model via the Peierls substitution, manifesting as a dimensionless Aharonov--Bohm phase $\phi$ accumulated by an electron when hopping between adjacent lattice sites. For a uniform magnetic field perpendicular to the plane of the ring, the vector potential $\mathbf{A}$ induces a phase factor between neighboring sites $j$ and $j+1$, defined by the line integral $\phi=\frac{e}{\hbar} \int_{j}^{j+1} \mathbf{A} \cdot d\mathbf{l}$. Applying Stokes' theorem to the entire ring and assuming that the flux is uniformly distributed over the $N$ bonds, the total magnetic flux $\Phi$ is related to the single-bond phase $\phi$ by $\Phi = N \phi$. For convenience, we introduce the dimensionless total flux $\widetilde{\Phi}_0 = \Phi / \Phi_0$, where $\Phi_0 = h/2e$ is the superconducting flux quantum. The single-bond phase can then be compactly expressed as $\phi = \pi \frac{\widetilde{\Phi}_0}{N}$.
 
 It should be emphasized that our model imposes specific physical constraints. Unlike systems with defects or weak links—where the magnetic flux penetrates only at the defect site, leading to localized phase accumulation \cite{Lucignano2013,Nava2017}—we consider a fully gapped, defect-free superconducting ring without any weak junctions. In this setting, the magnetic flux phase is uniformly distributed across all nearest-neighbor hopping terms: every hopping process acquires the same phase $\phi$. This treatment ensures a direct linear correspondence between the microscopic hopping parameter and the macroscopic magnetic flux, thereby enabling a more accurate simulation of topological properties in a homogeneous superconducting ring. Furthermore, the uniform distribution of the Peierls phase $\phi$ across all bonds ensures that the Hamiltonian retains translational invariance around the ring. This allows us to apply Bloch’s theorem and perform a Fourier transform to momentum space. The resulting Bogoliubov–de Gennes (BdG) Hamiltonian reads:
 \begin{eqnarray}
 	H(k) &=& 2t \sin k \sin \phi\sigma_0 -2|\Delta| \sin k \sin(\theta +\phi )\sigma_x \notag \\
 	     &-& 2 |\Delta| \sin k \cos (\theta+\phi ) \sigma_y +\big(- \mu-2t \cos k \cos \phi \big) \sigma_z ,\notag \\
 \end{eqnarray}
 where $k = 2\pi n / N$ ($n = 0, 1, ..., N-1$) are the discrete momenta, $\sigma_{y,z}$ are Pauli matrices acting in Nambu (particle-hole) space, and we have set the superconducting phase $\theta = 0$ without loss of generality.
 
 This Hamiltonian can be derived from Eq.~(1) by substituting the Fourier transforms:
 \[
 c_j = \frac{1}{\sqrt{N}} \sum_k e^{ikj} c_k, \quad c_j^\dagger = \frac{1}{\sqrt{N}} \sum_k e^{-ikj} c_k^\dagger,
 \]
 and using the identity $\sum_j e^{i(k-k')j} = N \delta_{k,k'}$. The resulting energy eigenvalues are given by
 \begin{equation}
 	E(k) = 2t\sin k \sin \phi \pm \sqrt{ \big( \mu + 2t \cos k \cos \phi \big)^2 + 4|\Delta|^2 \sin^2 k }.
 \end{equation}
 Note that the spectrum depends on the total flux through the combination $\phi$, and is independent of the superconducting phase $\theta$, as expected for a homogeneous ring.

 We employed the Landauer-B{\"u}ttiker formula in conjunction with the non-equilibrium Green's function approach to compute the transmission coefficients $T_\text{DT}$, $T_\text{LAR}$ and $T_\text{CAR}$ of different transport mechanisms, respectively \cite{Lee1981,Jauho1994}.\\
    \begin{eqnarray}
 T_\text{DT}&=\text{Tr}[\Gamma_{\text{Lee}} G_{\text{Cee}}^{\text{r}}\Gamma_{\text{Ree}}G_{\text{Cee}}^{\text{a}}],\\
 T_\text{LAR}&=\text{Tr}[\Gamma_{\text{Lee}} G_{\text{Ceh}}^{\text{r}}\Gamma_{\text{Lhh}}G_{\text{Ceh}}^{\text{a}}],\\
 T_\text{CAR}&=\text{Tr}[\Gamma_{\text{Lee}} G_{\text{Ceh}}^{\text{r}}\Gamma_{\text{Rhh}}G_{\text{Ceh}}^{\text{a}}],
   \label{Eq2}
   \end{eqnarray}
where $\text{Tr}$ denotes the matrix trace. The notation $\text{e/h}$ represents electrons (holes), while $\text{L/R}$ represent the left (right) electrodes. The subscript  "C" designates the central region. $\Gamma_{\text{L/R}}$ represents the left (right) broadening function,
where $\Gamma_{\text{L/R}}=i(\Sigma_{\text{L/R}}^{r}-\Sigma_{\text{L/R}}^{a})$, and $\Sigma_{\text{L/R}}$ represents the left (right) self-energy. $\Gamma_{\text{Lee}}$ means an electron enter in the left hand, $\Gamma_{\text{Ree}}$ means an electron is emitted from the right end, and $\Gamma_{\text{Lhh}}$ ($\Gamma_{\text{Rhh}}$) means is reflected to the left (right) end.
$G^{\text{r/a}}$ represents the retarded Green's function and the advanced Green's function respectively, and they satisfy the relation $G^{\text{r}}={G^{\text{a}}}^{\dagger}=(E-H-\Sigma_{\text{L}}^{\text{r}}-\Sigma_{\text{L}}^{\text{a}})^{-1}$.
$T_{\text{DT}}$, $T_\text{LAR}$ and $T_\text{CAR}$ correspond to microscopic processes of DT, LAR and CAR, respectively shown in Fig. \ref{Fig1}.

\begin{figure}[tb]
%\begin{center}
\centering
\includegraphics[width=0.45\textwidth]{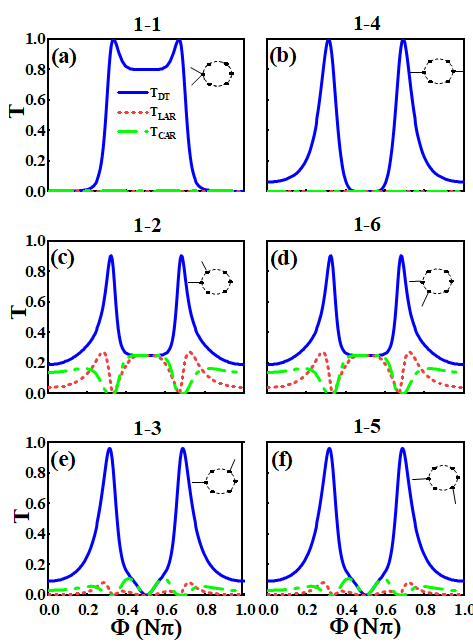}
%  \end{center}
\caption{ The transmission coefficients $T_\text{DT}$, $T_\text{LAR}$ and $T_\text{CAR}$ as functions of the magnetic flux phase and different connection configuration of the electrodes (insets in every subfigures). The ring has $N = 6$ lattice points. The on-site potential is $\mu = 1$, the hopping energy is $t = 1$, and the superconducting pairing potential is $\Delta = 1$ with $\theta = 0$. In (a)-(f), the left-hand electrode is fixed to connect to the lattice point $j = 1$ of the ring, while the right-hand electrode is connected to the lattice points $j = 1$, $j = 4$, $j = 2$, $j = 6$, $j = 3$, and $j = 5$ in sequence.}
\label{Fig2}
\end{figure}

\section{RESULTS AND DISCUSSION}\label{SC3}
Now, we focus on the impact of the parity (odd or even) of $N$ on the transport mechanisms in the Kitaev ring. It has been noted that the Kitaev chain is a topologically nontrivial phase for $\left| \mu \right|<2t$, and a topologically trivial phase for $\left| \mu \right|>2t$ in the absence of a magnetic field \cite{Kitaev2001}. Here, we set the parameters as follows: the hopping energy is $t = 1$, and the superconducting pairing potential is given by $\Delta = 1$ with a phase $\theta = 0$, the on-site potential is set to $\mu = 1$.

Since the model is in a ring shape, it satisfies rotational symmetry. As depicted in Fig. \ref{Fig2}, we take a ring composed of six lattice points as a representative example. The left-hand electrode is firmly fixed to the first lattice point of the chain, and subsequently, the right-hand electrode is systematically shifted from $j = 2$ to $j = 6$.
When the right-hand electrode is connected to lattice points $j = 1$ and $j = 4$, a symmetric configuration of the left and right electrodes is established. Under such symmetric conditions, all the transmitted particles are free electrons, and both local Andreev reflection and crossed Andreev reflection are effectively suppressed.
As the magnetic field phase changes, the direct transmission coefficient exhibits two resonant peaks in the vicinity of $N\pi/3$ and $2N\pi/3$, which arises from the system exhibiting two topological phase transitions at $\mu=1$, with the topologically trivial and nontrivial phases being separated by boundaries at $N\pi/3$ and $2N\pi/3$, the $Z_2$ topological invariant characterizing the phase is discussed in \cite{Miao2022}, see in particular Fig. 3(f) for a visual representation of its behavior across the topological phase transition.
Along with a distinct trough around $\Phi=N\pi/2$,
by referring to Eq. (\ref{Eq1}), it becomes evident that $\Phi$ exerts an influence on the hopping energies of electrons and holes. At $\Phi = N\pi/2$, the hopping energies of electrons and holes reach an equality, resulting in a state of dynamic equilibrium that gives rise to the observed trough.
When the electrodes are asymmetrically coupled to lattice of the ring, in addition to $T_\text{DT}$, $T_\text{LAR}$, and $T_\text{CAR}$ also manifest.
In the context of LAR, a single incoming electron being reflected as a hole, with a corresponding Cooper pair added to the superconductor, as shown in Fig. \ref{Fig1}(c).
Regarding CAR, a single electron impinges on the superconductor interface, a hole is reflected from the opposite interface, with a Cooper pair enters into the superconductor, as illustrated in Fig. \ref{Fig1}(d).
To facilitate a more intuitive observation of the results, we have reorganized the sequence of the figures. It is apparent that when the right-hand electrode is connected to lattice point $j = 2$, the $T - \Phi$ curve is indistinguishable from that when it is connected to lattice point $j = 6$.
This finding implies that there is no discernible difference between clockwise and counter-clockwise wiring configurations for these two cases.
This conclusion can also be corroborated in the cases of $j = 3$ and $j = 5$ in Fig. \ref{Fig2}(e) and Fig. \ref{Fig2}(f).
\begin{figure}[tb]
  %\begin{center}
  \centering
  \includegraphics[width=0.45\textwidth]{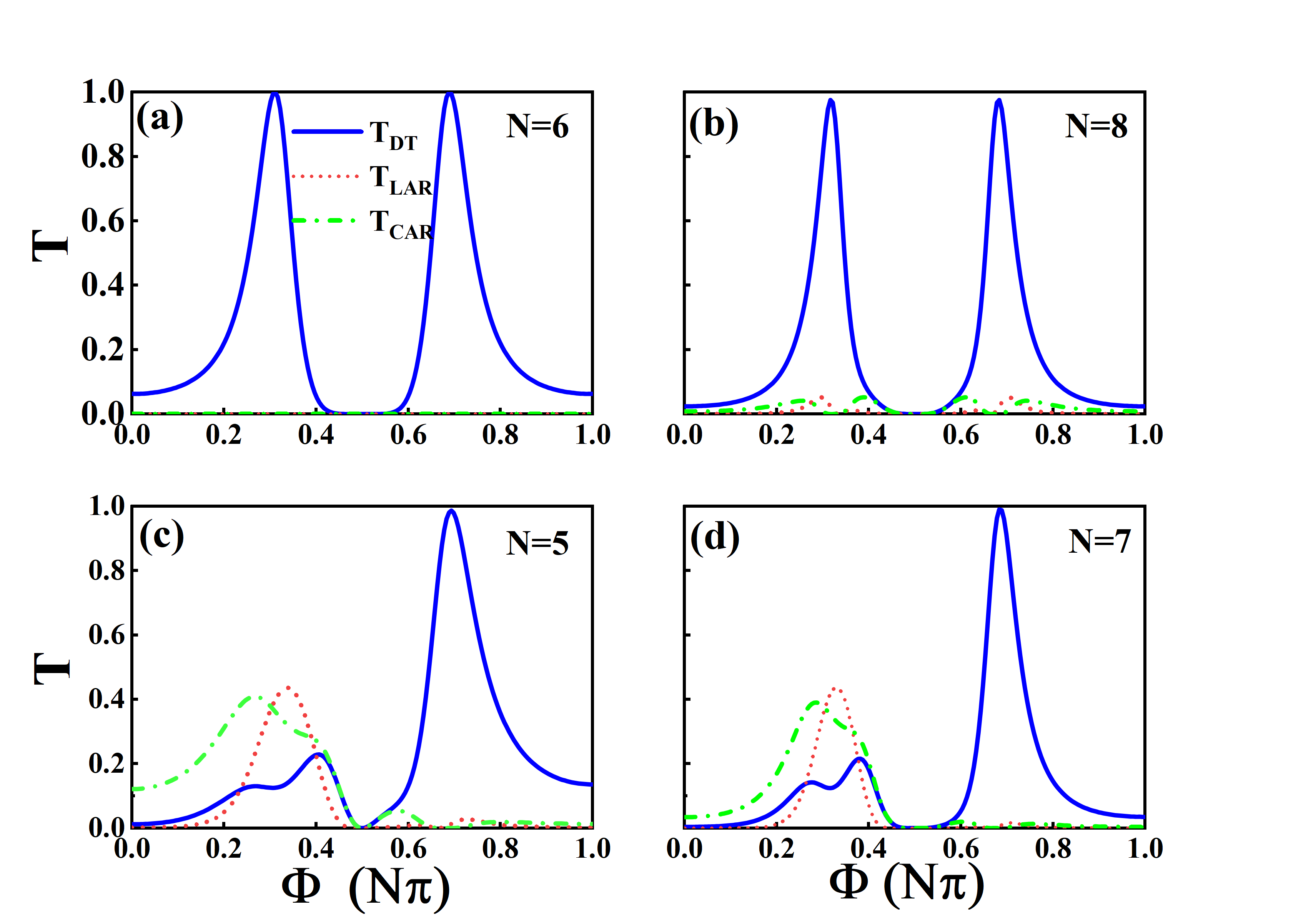}
  %\end{center}
  \caption{The transmission coefficients $T_\text{DT}$, $T_\text{LAR}$, and $T_\text{CAR}$ as functions of the magnetic flux phase and different lattice-point configurations. The left-hand electrode is connected to the lattice point $j = 1$, while the right-hand electrode is connected to the lattice point $j = 4$. The on-site potential is set to $\mu = 1$, the hopping energy is $t = 1$, and the superconducting pairing potential is given by $\Delta = 1$ with a phase $\theta = 0$. In (a)-(d), the number of lattice points in the system is set to $N = 6$, $N = 8$, $N = 5$, and $N = 7$ successively. }
  \label{Fig3}
  \end{figure}

In Fig. \ref{Fig2}, we also show how the transmission coefficient varies with the connection configurations of the electrodes to the ring for even $N$. We want to emphasize that the DT is always dominant and the LAR and CAR are weaker than the DT process due to a symmetric ring (even $N$).
In the case of even $N$ and symmetric connection of electrodes to the ring, $T_\text{LAR}$ and $T_\text{CAR}$ are even almost vanishing [Fig. \ref{Fig2}(b)]. These behaviors of transmission through a Kitaev ring reflects a competition among these three mechanisms. What we find in this work is that things are changed dramatically for an asymmetric connection of electrodes even for even $N$ case [see Figs. \ref{Fig2}(c) and \ref{Fig2}(d)], in which $T_\text{LAR}$ and $T_\text{CAR}$ develop finite values due to the breaking of the symmetry connection of electrodes. It can be understood that the single electron transport in this case is somehow reduced and $T_\text{LAR}$, and $T_\text{CAR}$transport processes are enhanced.

In this work, another finding is that qualitatively distinct behavior of transmission for odd $N$ (we call it as asymmetric ring) exists. To study this case, we arrange the left electrode coupled to $j = 1$ site, and the right electrode to $j = 4$ site of the Kitaev ring.
We plot $T_\text{DT}$, $T_\text{LAR}$, and $T_\text{CAR}$ with the magnetic flux phase $\Phi$ for different $N$.
For $N = 6$ [Fig. \ref{Fig3}(a)], $T_\text{DT}$ dominates the transport, and exhibits two resonant peaks at $N \pi/3$ and $2N \pi/3$, which are extremely remarkably symmetric with respect to $\Phi=N \pi/2$.
For the case of $N = 8$ [Fig. \ref{Fig3}(b)], the transport characteristic curve is similar to $N = 6$.
However, it differs from the $N = 6$ scenario in that a small but observable number of Cooper pairs exist. The nonzero $T_\text{LAR}$ and $T_\text{CAR}$ can be attributed to the asymmetry between the left and right electrodes (Fig. \ref{Fig2}).
When $N = 5$ and $N = 7$ [Figs. \ref{Fig3}(c) and \ref{Fig3}(d)], the transport coefficient ($T_\text{DT}$, $T_\text{LAR}$ and $T_\text{CAR}$) are asymmetric with respect to $\Phi=N \pi /2$. That means the peak of $T_\text{DT}$ around $N \pi/3$ is largely suppressed, while the peak at $2N \pi/3$ almost the same.
For a comparison, almost vanishing $T_\text{LAR}$ and $T_\text{CAR}$ for even $N$ case develop prominent and broad peaks around $N\pi/3$ for odd $N$ case.
%
%$T_\text{DT}$ has a peak near the magnetic flux phase of $2N \pi/3$, while near the magnetic flux phase $N \pi/3$, the values of $T_\text{LAR}$ and $T_\text{CAR}$ are relatively large in which case the particles transported within the ring include not only electrons but also Cooper pairs.
%
As the magnetic flux phase increases, the values and variations of these peaks are highly complex and do not follow an obvious symmetry rule.
In conclusion, it can be deduced that the symmetry property of the transport characteristic curve with respect to $\Phi=N \pi/2$ depends on whether $N$ in the ring is odd or even.
\begin{figure}[tb]
	\centering
	\includegraphics[width=0.45\textwidth]{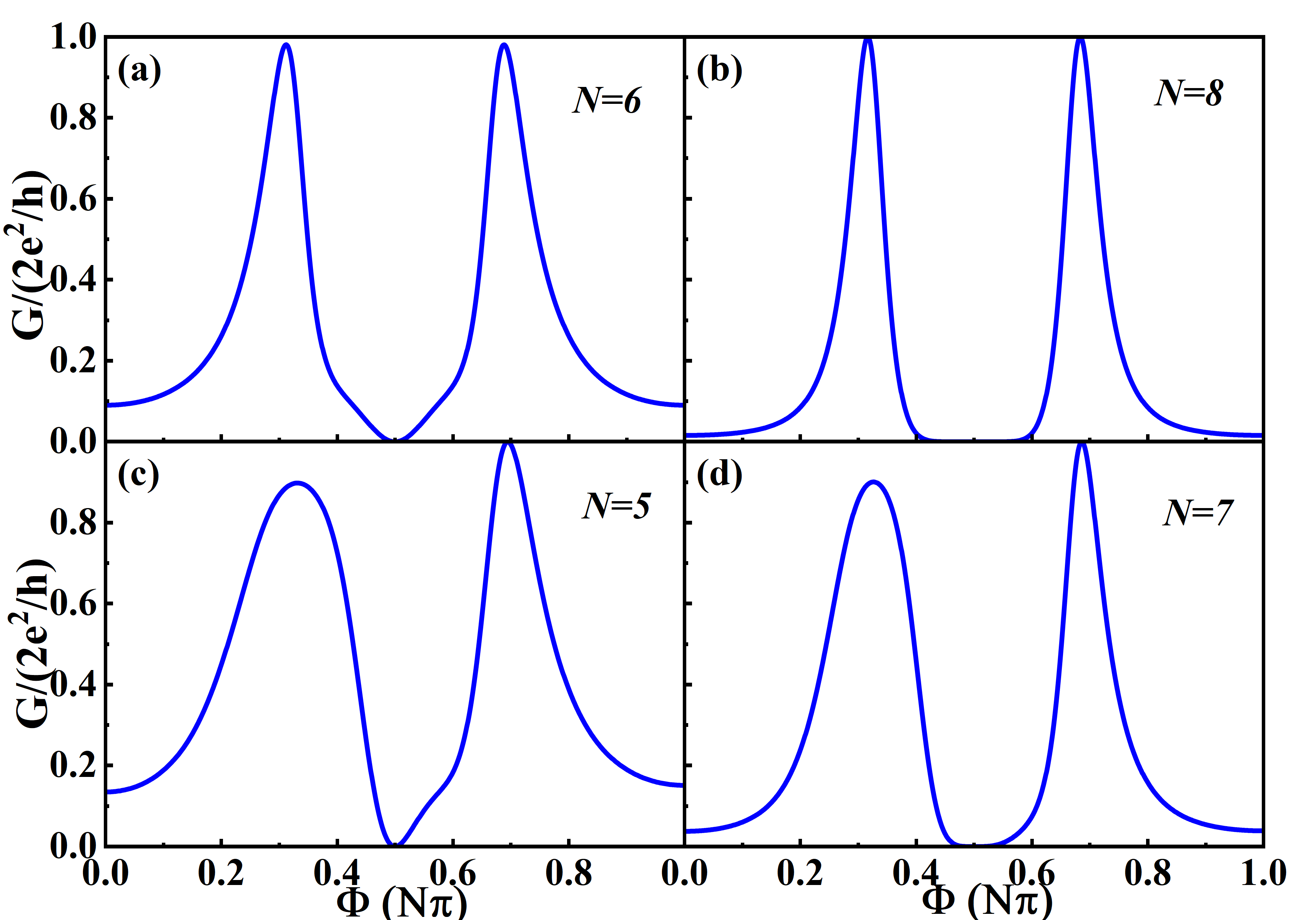}
	\caption{ The differential conductance $G$ as functions of the magnetic flux phase and different lattice-point configurations. The on-site potential is set to $\mu = 1$, the hopping energy is $t = 1$, and the superconducting pairing potential is given by $\Delta = 1$ with a phase $\theta = 0$. In (a)-(d), the number of lattice points in the system is set to $N = 6$, $N = 8$, $N = 5$, and $N = 7$ successively.}
	\label{Fig4}
\end{figure}

 To better connect with the experiments, we calculated the differential conductance $ G = dI/dV $ using the Landauer-B{\"u}ttiker formula, and present in Fig. \ref{Fig4} a clear illustration of the dependence of the differential conductance on the magnetic flux. Figs. \ref{Fig4}(a) and (b) show that for even $ N $, the $ G $-$\Phi $ curves are symmetric about $ \Phi = N\pi/2 $, with peak values approaching 1. In contrast, for odd $ N $, the $ G $-$\Phi $ curves exhibit asymmetry about $ \Phi = N\pi/2 $, and the peak values at $ \Phi = N\pi/3 $ are smaller than 1. These features are consistent with the conclusions drawn from the $ T $-$\Phi $ curves.

\begin{figure}[tb]
	%\begin{center}
	\centering
	\includegraphics[width=0.5\textwidth]{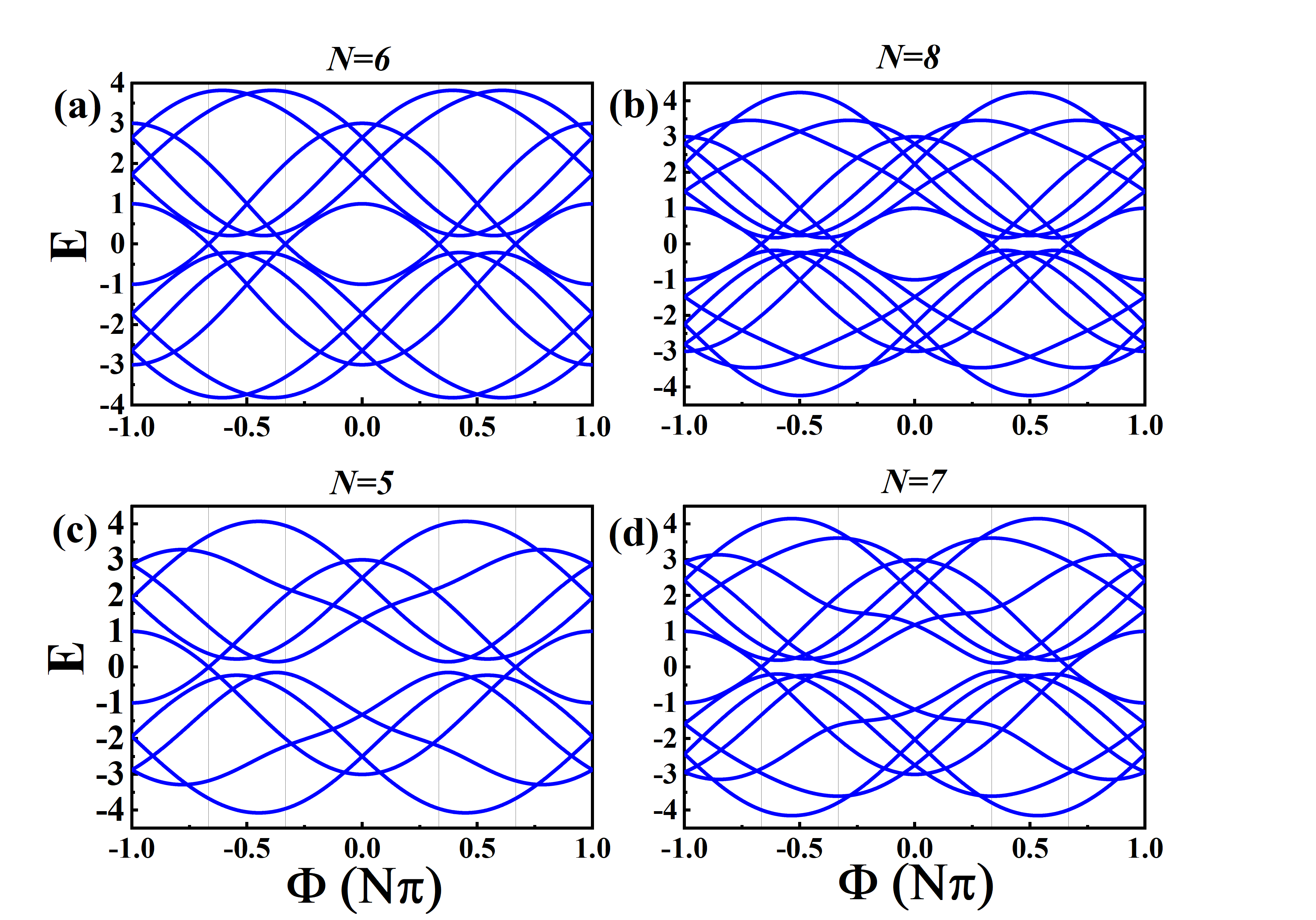}
	%\end{center}
	\caption{The plots show the energy ($E$) as a function of the magnetic flux phase ($\Phi$) for different numbers of lattice points, presented as $E - \Phi$ diagrams. In (a)-(d), $N$ in the ring is set to $N = 6$, $N = 8$, $N = 5$, and $N = 7$ respectively. }
	\label{Fig5}
\end{figure}

To gain a deeper understanding of the phenomenon where the $T - \Phi$ plots display distinct central symmetries for odd and even numbered lattice points in the ring, we proceed to conduct a detailed exploration of the $E- \Phi$ relationship diagram.
As clearly depicted in Figs. \ref{Fig5}(a) and \ref{Fig5}(b), where $N$ in the ring is even, for instance $N = 6$ and $N = 8$, the energy gap closes precisely at $\Phi=N\pi/3$ and $\Phi=2N\pi/3$ which correspond the location of two resonant peaks in Figs. \ref{Fig3}(a) and \ref{Fig3}(b).
At these points, the dominant species of transported particles within the ring are free electrons.
In contrast, when $N$ within the ring is odd, for instance $N = 5$ and $N = 7$ [Figs. \ref{Fig5}(c) and \ref{Fig5}(d)], the energy gap keeps opening in the $E - \Phi$ diagram in the vicinity of $\Phi=\pm N \pi/3$.
As clearly depicted in Figs. \ref{Fig3}(c) and \ref{Fig3}(d), the local Andreev reflection coefficients $T_\text{LAR}$ and the crossed Andreev reflection coefficient $T_\text{CAR}$ exhibit peaks and become dominant around $\Phi=N \pi/3$.
Owing to the tunneling effect, free electrons retain a non-negligible probability of traversing the energy gap. When $\Phi=\pm 2N \pi/3$, the energy gap in the $E - \Phi$ plot closes, indicating that the particles transported within the ring are free electrons.
Upon comparing Figs. \ref{Fig5}(a) and \ref{Fig5}(c), it becomes evident that in the case of even $N$, the $E - \Phi$ plot exhibits symmetry with respect to $\Phi = N \pi/2$ and has a period of $N \pi$.
Conversely, when $N$ is an odd, the $E - \Phi$ plot is symmetric about $\Phi=0$, and its period is $2N \pi$.
The distinct differences in the symmetry and periodicity characteristics of the $E - \Phi$ graphs for odd and even values of $N$ provide crucial clues regarding the different symmetries corresponding to the odd and even numbers of lattice points within the ring.

\begin{figure}[tb]
\centering
\includegraphics[width=0.48\textwidth]{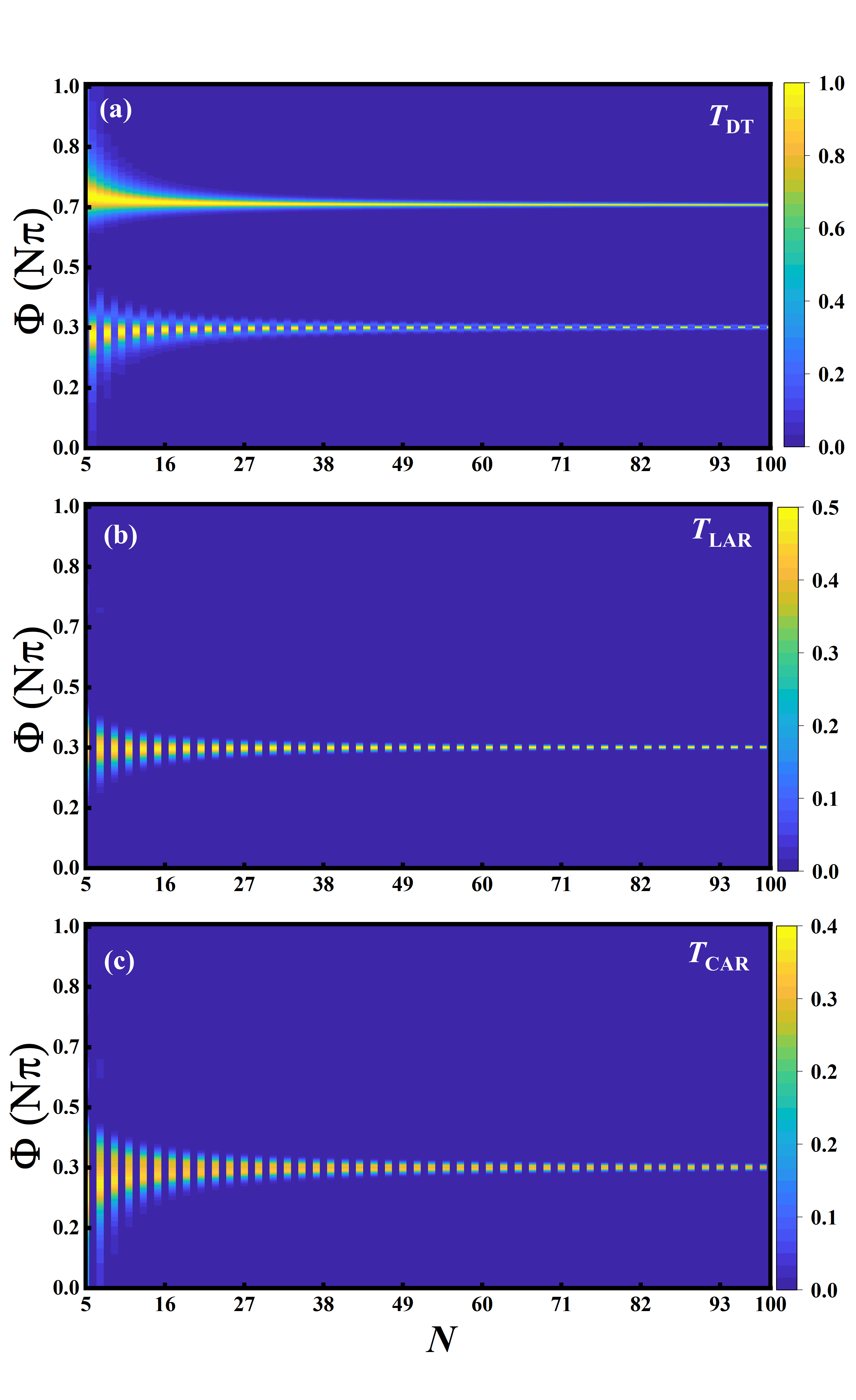}
\caption{The plots depict the variations of the transmission coefficients: (a) $T_\text{DT}$, (b) $T_\text{LAR}$, and (c) $T_\text{CAR}$, as functions of $\Phi$ and $N$. The yellow regions in the plots represent the peak values, whereas the dark-blue regions indicate zero values. }
\label{Fig6}
\end{figure}

To validate the generality of our findings, we systematically expand the radius of the ring and generate heatmaps to illustrate how the transmission coefficients vary as functions of $N$ and the magnetic flux $\Phi$.
In Fig. \ref{Fig6}, as $N$ increases, all transmission coefficients exhibit periodic oscillations.
For a ring with even $N$, at $\Phi = N \pi/3$, the close of superconducting energy gap [Figs. \ref{Fig5}(a) and \ref{Fig5}(b)] allows free electrons to pass through which leads to the maximum of $T_\text{DT}$ and the vanishing of $T_\text{LAR}$ and $T_\text{CAR}$.
In contrast, for a ring composed of an odd $N$, when $\Phi = N \pi/3$, a superconducting energy gap emerges [Figs. \ref{Fig5}(c) and \ref{Fig5}(d)] which means free electrons could not to pass through.
Consequently, this gives rise to pronounced contributions from $T_\text{LAR}$ [as depicted in Fig. \ref{Fig1}(c)] and $T_\text{CAR}$ [as shown in Fig. \ref{Fig1}(d)].
As increasing of $N$, the regions associated with the maximum values of the coefficients become progressively narrower but remain present. This persistence confirms the robustness and universality of our conclusions across varying ring sizes.
To summarize briefly, we observe there is always a peak located around $\Phi=2N \pi/3$, while the peak at $\Phi=N \pi/3$ alternately appears and disappears for $T_\text{DT}$. For $T_\text{LAR}$ and $T_\text{CAR}$, the peaks from the contribution of $\Phi=2N \pi/3$ disappear completely, and those at $\Phi=N \pi/3$ exhibit appearance-disappearance oscillation for odd-even $N$. This distinct feature for the transport through symmetric and asymmetric Kitaev ring is an inherent character for even-odd lattice number ring.

\begin{figure}[tb]
	\centering
	\includegraphics[width=0.48\textwidth]{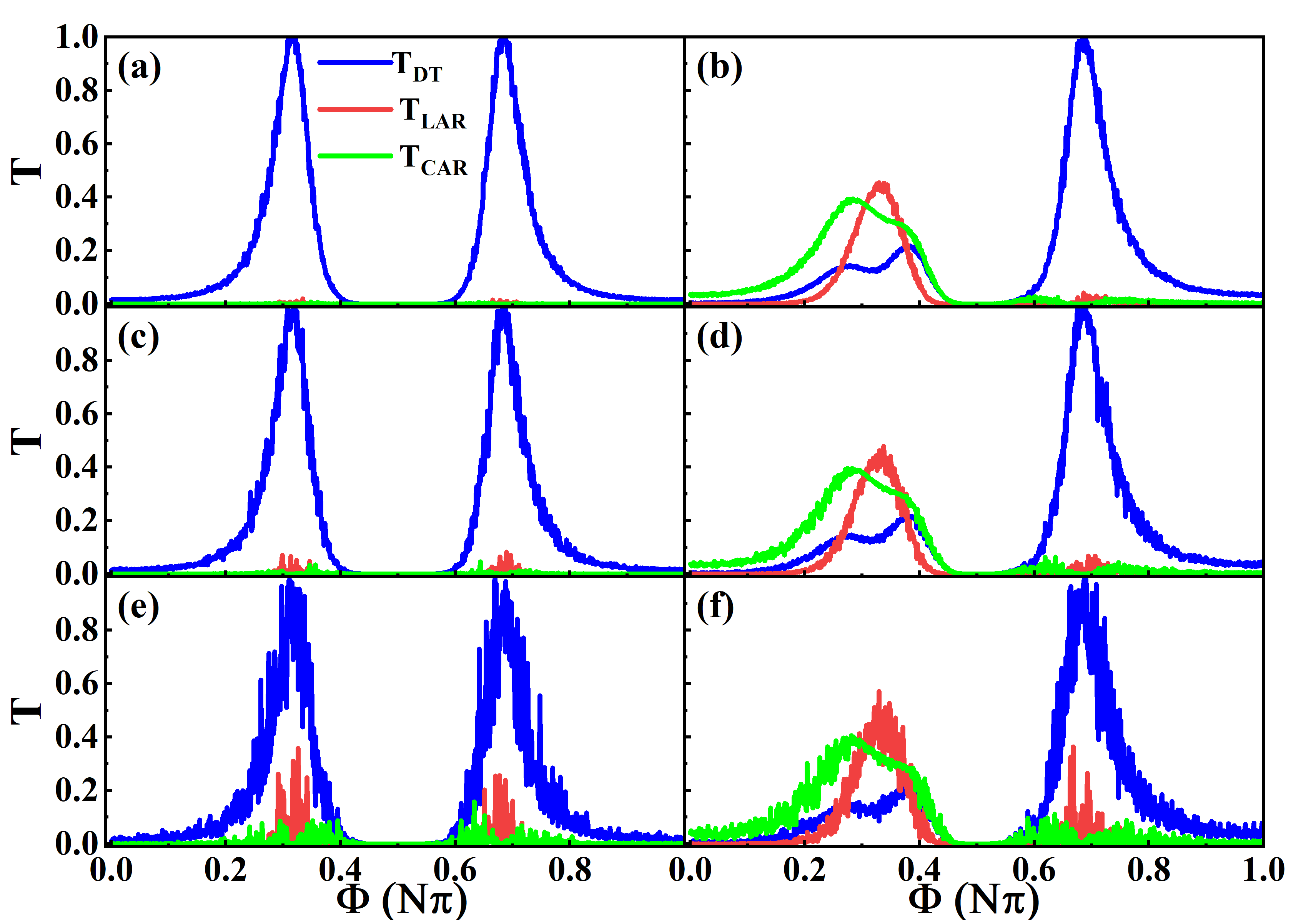}
	\caption{The transmission coefficients $T_\text{DT}$, $T_\text{LAR}$ and $T_\text{CAR}$ as functions of the magnetic flux phase and different disorder strength W. $N=8$ for the left three panels, while $N=7$ for the right three panels Figures. (a-b) are for a disorder strength of $\text{W}=0.1$, (c-d) for $\text{W}=0.2$, and (e-f) for $\text{W}=0.5$.}
	\label{Fig7}
\end{figure}

 To demonstrate that the parity-dependent effect is not merely an artifact of the idealized clean limit, but rather exhibits a degree of robustness rooted in the underlying symmetry of the system, we have carried out a systematic analysis of the impact of weak disorder on the transport properties. Under these conditions, the system Hamiltonian becomes:
\begin{eqnarray}
	%  \begin{aligned}
		H_\mathrm{C} &=& \sum_{j=1}^{N}-\mu_{j} c_{j}^{\dagger} c_{j} + \sum_{j=1}^{N-1}\left(-t e^{i\phi} c_{j}^{\dagger} c_{j+1}+|\Delta| e^{i\theta} c_{j}^{\dagger}c_{j+1}^{\dagger}\right) \notag\\
		& + & \left(-t e^{i\phi} c_{N}^{\dagger} c_{1}+|\Delta|e^{i\theta} c_{N}^{\dagger}c_{1}^{\dagger}\right)+\text{H. C.},
		\label{Eq5}
		%  \end{aligned}
\end{eqnarray}
where $\mu_{j}=\mu+w_j$, with $w_j$ is a random number uniformly distributed in the interval $[-W,W]$. In general, topological properties may be altered in the presence of strong disorder. For clarity of discussion, we restrict ourselves to the weak-disorder regime. Fig. \ref{Fig7} shows that, as the disorder strength increases, $T_\text{LAR}$ and $T_\text{CAR}$ can be induced at $ N\pi/3 $ for even $ N $. The oscillations in the conductance curves become more pronounced with increasing disorder, yet the positions of the peaks and their symmetry remain unchanged. Unlike the teleportation effect \cite{Milburn1991,Oh2002,Cacciapuoti2020,Hosseiny2024}-where MZMs undergo non local coupling through a topological superconducting ring, giving rise to a 4$\pi$-periodic Josephson current-which signature is difficult to observe experimentally due to decoherence and finite-size effects, our work highlights an alternative, parity-sensitive transport signature that persists even in the presence of moderate disorder. This provides a potentially more accessible approach to probing the topological nature of finite-sized rings. Therefore, we conclude that the system retains its underlying symmetry in the presence of weak disorder.

\section{SUMMARY AND OUTLOOK}\label{SC4}
In this paper, we study the variation of the transport coefficients in an annular Kitaev chain as a function of the magnetic field and the total lattice number of Kitaev ring. %
By adjusting $N$ in the ring, we find that when $N$ is odd, the $T - \Phi$ diagram is not symmetric about $\Phi=N \pi/2$, while $T - \Phi$ diagram is symmetric for even $N$.
To be specific, there are two transmission peaks of $T_\text{DT}$ for even $N$ which has been predicted previously, $T_\text{LAR}$ and $T_\text{CAR}$ are suppressed. We find in this work that in even $N$ case, $T_\text{LAR}$ and $T_\text{CAR}$ developed finite values when the electrodes are asymmetrically connected to the ring.
Another observation is that the two peaks of $T_\text{DT}$ located at $\Phi=N \pi/3$ and $2N \pi/3$ becomes asymmetric for odd $N$ Kitaev ring.
There emerge peaks for $T_\text{LAR}$ and $T_\text{CAR}$ at $\Phi=N \pi/3$, while no peak exists around $2N \pi/3$. In fact, it is observed that $T_\text{LAR}$ and $T_\text{CAR}$ are quite suppressed around $2N \pi/3$.
These observations may be understood by the gap opening or closing around $\Phi=N \pi/3$ and $\Phi=2N \pi/3$. It is always gapless at $2N \pi/3$ which may be the reason for the disappearance of $T_\text{LAR}$ and $T_\text{CAR}$ and a robust peak of $T_\text{DT}$.
Moreover, the energy gap at $\Phi=N \pi/3$ is opened and closed with varying $N$ from odd to even. As a consequence, $T_\text{LAR}$ and $T_\text{CAR}$ emerge peaks for odd $N$ but disappear for even $N$ (symmetric connection of electrodes). As can be seen from our results, near the magnetic flux phase $\Phi=N \pi/3$, in the case $N$ is an even number, DT is the most optimal; while in the case $N$ is an odd number, LAR and CAR are the most optimal.
We conclude that the parity of Kitaev ring lattice number dramatically determine the quantum transport through such a ring system.
%
%It is expected to provide solid theoretical support for the experimental design of observable parity-dependent transport characteristics.
%
  In view of recent advances in semiconductor–superconductor quantum-dot arrays \cite{Bordin2024,Bordin2025,Wang2022,Dvir2023}, the parity-dependent transport signatures predicted here may be accessible in near-term experiments.
Thus we expect that the parity-dependent transport characteristics might be observed in future experiments.

\begin{acknowledgments}
This work is supported in part by National Key R\&D Program of China (Grant No. 2024YFA1409200, and No. 2022YFA1402802),  the Training Program of Major Research plan of NSFC (Grant No. 92165105), and CAS Project for Young Scientists in Basic Research Grant No. YSBR-057. G. S. was supported in part by the Quantum Science and Technology-National Science and Technology Major Project under Grant No. 2024ZD0300500, NSFC Nos. 12534009 and 12447101, the Strategic Priority Research Program of Chinese Academy of Sciences (Grant No. XDB1270000) and the CAS Superconducting Research Project under Grant No. SCZX-0101.
\end{acknowledgments}

\appendix
\setcounter{figure}{0}
\renewcommand{\thefigure}{A\arabic{figure}}
\renewcommand{\thetable}{A\arabic{table}}
%\begin{widetext}
\section{TRANSMISSION COEFFICIENT} % the title of appendix
\label{AppendixA}
Here we introduce the specific calculation method of the transmission coefficient in this paper. We split the total Hamiltonian in three pieces, $H=H_\text{C}+H_\text{W}+H_\text{T}$, where $H_\text{C}$ modes the central region, $H_\text{W}$ describes the contacts and $H_\text{T}$ is the the tunneling coupling between contacts and the central region. Below we discuss each of these terms. 

The central region Hamiltonian is $H_\text{C}=\psi^\dagger h_\text{c} \psi$, while $h_\text{c}$ is the matrix in the basis $\psi^\dagger=(c_1^\dagger,c_1, c_2^\dagger, c_2, \dots,c_N^\dagger,c_N)$. From Eq.(1) in the maintex, the tight-binding Hamiltonian  in the real-space of the central region is:
\begin{equation}
	h_\text{c}=\begin{pmatrix}
		h_{11} & h_{12}  & 0 & 0  &\dots & 0 &h_{1N}\\
		h_{21}& h_{22} & h_{23} & 0 &\dots & 0 &0\\
		0& h_{32} & h_{33} & h_{34} & \dots & 0 &0\\
		\vdots & \vdots & \ddots & \ddots & \ddots &\vdots &\vdots \\
		\vdots & \vdots & \vdots & \ddots & \ddots &\ddots &\vdots \\
		0& 0 & 0 & \dots & h_{N-1 N-2} & h_{N-1 N-1} &h_{N-1 N}\\
		h_{N1}& 0& 0  & \dots & 0 &h_{N N-1} &  h_{N N}
		\label{Eq5}
	\end{pmatrix},
\end{equation}
$ h_{j,j}$ denotes the energy associated with the $j $ lattice site in real-space, while $ h_{j,j+1} $ and $ h_{j-1,j} $ represents the interaction term between neighboring lattice sites. Diagonalize it using the $u$-matrix transformation. The $u$-matrix satisfy the condition of $uu^\dagger=1$. Then we have $H_c=\psi^\dagger u^\dagger u h_\text{c} u^\dagger u \psi$. Define the $\Psi^\dagger=\psi^\dagger u^\dagger$, $\Psi^\dagger=(d_1^\dagger, d_2^\dagger,\dots,d_N^\dagger)$.
\begin{equation}
	u h_\text{c} u^\dagger =\begin{pmatrix}
		\epsilon_{1} & 0  & 0 & 0 & \dots & 0 \\
		0& \epsilon_{2} & 0 & 0 & \dots & 0 \\
		0& 0 & \epsilon_{3} & 0 & \dots & 0 \\
		\vdots & \vdots & \vdots & \ddots & \vdots& \vdots \\
		0& 0 & \vdots & 0 & \epsilon_{N-1} &0 \\
		0& \vdots & 0 & 0 & 0 &  \epsilon_{N}
		\label{Eq6}
	\end{pmatrix}
	,
\end{equation}
\begin{eqnarray}
	H_\text{C}=\sum_n \epsilon_n d_n^{\dagger} d_n.
	\label{Eq7}
\end{eqnarray}

In our work, we think the electrons in the leads as noninteracting except for an overall self-consistent potential. Thus, the contact Hamiltonian is
\begin{eqnarray}
	H_\text{W}=\sum_{k,w\in \text{L/R} } \epsilon_{kw} c_{kw}^{\dagger}c_{kw},
	\label{Eq8}
\end{eqnarray}
and the Green functions in the leads for the uncoupled system are:
\begin{eqnarray}
	g_{kw}^{<}(t-t')&=&i\left \langle c_{kw}^{\dagger}(t') c_{kw}(t) \right \rangle \notag\\
	&=&if(\epsilon_{kw}^{0}) \text{exp}[-i \epsilon_{kw} (t-t')],
	\label{Eq9}
\end{eqnarray}
\begin{eqnarray}
	g_{kw}^{\text{r,a}}(t-t')&=&\mp i\theta (\pm t \mp t')\left \langle \left \{ c_{kw}(t), c_{kw}^{\dagger}(t') \right \}  \right \rangle \notag\\
	&=&\mp i\theta (\pm t \mp t') \text{exp}[-i \epsilon_{kw} (t-t')],
	\label{Eq10}
\end{eqnarray}
here $f(\epsilon_{kw})=[\text{exp}[(\epsilon_{kw}-\mu_w)/k_BT]+1]^{-1}$ is the equilibrium distribution in a given lead.

We assume the coupling constants between the leads and the central as known, and write
\begin{eqnarray}
	H_\text{T}=\sum_{k,m,w\in \text{L,R}}V_{kw,m}c_{kw}^{\dagger}d_m+\text{H.c.},
	\label{Eq11}
\end{eqnarray}
$d_m^{\dagger}(d_m)$ creates (destroys) an electron in state $\left | m  \right \rangle$, $m$ is the site on the ring which connect the electrode.
Consider the current situation at one end. Take the left end as an example.
\begin{eqnarray}
	J_{\text{L}}=-e \langle\dot{N_\text{L}} \rangle=-\frac{ie}{\hbar} \langle[H,N_\text{L}]\rangle,
	\label{Eq12}
\end{eqnarray}
$N_\text{L}$ represents the number of particles in the left electrode.
\begin{eqnarray}
	J_\text{L}=&-&\frac{ie}{\hbar} \langle[H,N_\text{L}]\rangle\notag\\
	=&-&\frac{ie}{\hbar} \left\langle \left[\sum_{k,m,w\in \text{L}}(V_{kw,m}c_{kw}^{\dagger}d_m +V_{kw,m}^*d_m^{\dagger}c_{kw}) \right.\right.\notag\\
	&+&\left.\left. \sum_n \epsilon_n d_n^{\dagger}d_n+\sum_{k,w\in \text{L}} \epsilon_{kw}c_{kw}^{\dagger} c_{kw},N_\text{L}\right] \right\rangle\notag\\
	=&-&\frac{ie}{\hbar} \sum_{k,m,w\in \text{L}} (V_{kw,m}^*\langle d_m^{\dagger}c_{kw}\rangle-H.C.),  
	\label{Eq13}
\end{eqnarray}
the less-than Green's function defined as varying with time. $G_{m,kw}^{<}(t-t')\equiv i\left \langle  c_{kw}^{\dagger}(t')d_m(t)\right \rangle $, $G_{kw,m}^{<}(t-t')\equiv i\left \langle  d_{m}^{\dagger}(t')c_{kw}(t)\right \rangle$,
and $G_{kw,m}^{<}(t,t)=-[G_{m,kw}^{<}(t,t)]^{*}$, then the current situation is:
\begin{eqnarray}
	J_\text{L}=\frac{2e}{\hbar} \Re \sum_{m,k,w\in \text{L}} V_{kw,m} G_{m,kw}^{<}(t,t),
	\label{Eq14}
\end{eqnarray}
\begin{eqnarray}
	& G_{m,kw}^{<} &(t-t') = \sum_{n,k,w\in \text{L}} \int dt_1 V_{kw,n}^{*}
	\left [G_{mn}^{\text{r}}(t-t_1) \right. \notag\\
	&\times & g_{kw}^{<}(t_1-t')+\left. G_{mn}^{<}(t-t_1)g_{kw}^{\text{a}}(t_1-t')\right],  %\notag\\
	\label{Eq15}
\end{eqnarray}
perform Fourier transform on the Eq. (\ref{Eq15}) ,
\begin{eqnarray}
	G_{m,kw}^{<}(\epsilon)= \sum_{n}&V_{kw,n}^{*} [G_{mn}^{\text{r}}(\epsilon)g_{kw}^{<}(\epsilon)  \notag\\
	& +G_{mn}^{<}(\epsilon)g_{kw}^{\text{a}}(\epsilon)],
	\label{Eq16}
\end{eqnarray}
then
\begin{eqnarray}
	J_\text{L} = \frac{2e}{\hbar}\int \frac{d\epsilon}{2\pi} &\Re& \sum_{n,m,k,w\in \text{L}}V_{kw,m} V_{kw,n}^{*}\left[G_{mn}^{\text{r}}(\epsilon)g_{kw}^{<}(\epsilon) \right.  \notag\\
	&+&  \left. G_{mn}^{<}(\epsilon)g_{kw}^{\text{a}}(\epsilon)\right], % \right\},
	\label{Eq17}
\end{eqnarray}
define broadening function $\Gamma^\text{L}{(\epsilon_k)}$, and $\rho(\epsilon_k)$ is the density of states.
\begin{eqnarray}
	[\Gamma^{\text{L}}{(\epsilon_k)}]_{nm}= 2\pi \sum_{w \in \text{L}}\rho_w(\epsilon_k)V_{w,m}(\epsilon_k) V_{w,n}^{*}(\epsilon_k),  %\notag\\
	\label{Eq18}
\end{eqnarray}
the Eq. (\ref{Eq17}) becomes
\begin{eqnarray}
	J_\text{L}=\frac{i e}{\hbar} \int \frac{d\epsilon}{2\pi}\text{Tr}( \Gamma^\text{L} (\epsilon) \{G^{<}(\epsilon)+f_\text{L}(\epsilon)[G^{\text{r}}(\epsilon)-G^{\text{a}}(\epsilon)]\}).\notag\\
	\label{Eq19}
\end{eqnarray}
In steady state, the current will be uniform and the total current have the following relationship.
\begin{eqnarray}
	J=J_\text{L}=-J_\text{R},\\
	J=\frac{J_\text{L}-J_\text{R}}{2},
	\label{Eq20}
\end{eqnarray}
and $D(\epsilon)=G^\text{r}(\epsilon)-G^\text{a}(\epsilon)$ is defined, thus, substituting Eq.(\ref{Eq19}) into Eq.(\ref{Eq20}) can obtain
\begin{eqnarray}
	J&=&\frac{ie}{2\hbar}\int \frac{d\epsilon}{2\pi} \text{Tr}\{[\Gamma^\text{L} (\epsilon)-\Gamma^\text{R} (\epsilon)] G^{<}(\epsilon)\notag\\
	&+&[f_\text{L}(\epsilon)\Gamma^\text{L} (\epsilon)-f_\text{R}(\epsilon)\Gamma^\text{R} (\epsilon)]D(\epsilon)\},  %\notag\\
	\label{Eq21}
\end{eqnarray}
introduce the left self-energy $\Sigma^{<}(\epsilon)$, and it satisfies the relationship $\Sigma^{<}(\epsilon)=i [f_\text{L}(\epsilon)\Gamma^\text{L} (\epsilon)+f_\text{R}(\epsilon)\Gamma^\text{R} (\epsilon)]$, and
$G^{<}(\epsilon)=G^{\text{r}}(\epsilon)\Sigma^{<}(\epsilon)G^{\text{a}}(\epsilon)$.
Substituting this relationship into Eq.(\ref{Eq21}) and yields,
\begin{eqnarray}
	J&=&\frac{ie}{2\hbar}\int \frac{d\epsilon}{2\pi} \text{Tr}\{ [\Gamma^\text{L} (\epsilon)-\Gamma^\text{R} (\epsilon)] G^{\text{r}}(\epsilon)\Sigma^{<}(\epsilon)G^{\text{a}}(\epsilon)\notag\\
	&+&[f_\text{L}(\epsilon)\Gamma^\text{L} (\epsilon)-f_\text{R}(\epsilon)\Gamma^\text{R} (\epsilon)]D(\epsilon)\},  %\notag\\
	\label{Eq22}
\end{eqnarray}
\begin{eqnarray}
	J&=&\frac{ie}{2\hbar}\int \frac{d\epsilon}{2\pi} \text{Tr}\{[f_\text{L}(\epsilon)-f_\text{R}(\epsilon)][i\Gamma^\text{R}(\epsilon) G^\text{r}(\epsilon) \Gamma^\text{L}(\epsilon) G^\text{a}(\epsilon)]\notag\\
	&+&f_\text{L}(\epsilon)[i\Gamma^\text{L}(\epsilon) G^\text{r}(\epsilon) \Gamma^\text{L}(\epsilon) G^\text{a}(\epsilon) + \Gamma^\text{L}(\epsilon) D(\epsilon)]\notag\\
	&-&f_\text{R}(\epsilon)[i\Gamma^\text{R}(\epsilon) G^\text{r}(\epsilon) \Gamma^\text{R}(\epsilon) G^\text{a}(\epsilon) + \Gamma^\text{R}(\epsilon) D(\epsilon)]\}, %\notag\\
	\label{Eq23}
\end{eqnarray}
utilize the Keldysh equation,
\begin{eqnarray}
	&D(\epsilon)=G^\text{r}(\epsilon)[\Sigma^{\text{r}}(\epsilon)-\Sigma^{\text{a}}(\epsilon)]G^\text{a}(\epsilon),\notag\\
	&\Sigma^{\text{r}}(\epsilon)-\Sigma^{\text{a}}(\epsilon)=-i\Gamma(\epsilon).
	\label{Eq24}
\end{eqnarray}
Substituting this relationship into Eq.(\ref{Eq23}) and yields,
\begin{eqnarray}
	J=\frac{e}{2\hbar}\int \frac{d\epsilon}{2\pi} \text{Tr}\{ [f_\text{L}(\epsilon)-f_\text{R}(\epsilon)][\Gamma^\text{L}(\epsilon) G^\text{r}(\epsilon) \Gamma^\text{R}(\epsilon) G^\text{a}(\epsilon)]\},\notag\\
	\label{Eq25}
\end{eqnarray}
which results that
\begin{eqnarray}
	T=\text{Tr}[\Gamma^\text{L}(\epsilon) G^\text{r}(\epsilon) \Gamma^\text{R}(\epsilon) G^\text{a}(\epsilon)].
	\label{Eq26}
\end{eqnarray}

%\newpage
%\bibliographystyle{apsrev4-2}
%\bibliography{reff}

\end{document}